\documentclass[12pt]{iopart}

\usepackage{graphicx}
\usepackage{epsfig}
\usepackage{amsbsy}
\usepackage{color}

\begin{document}
\title[Charge density mapping few-electron two-dimensional quantum dots]{Charge density mapping of strongly-correlated few-electron two-dimensional quantum dots by scanning probe technique}

\author{E. Wach, D.P. \.Zebrowski, and B. Szafran}
 \address{AGH University of Science and Technology, \\ Faculty of Physics and Applied Computer Science,
al. Mickiewicza 30, 30-059 Krak\'ow, Poland}

\date{\today}

\begin{abstract}
We perform a numerical simulation of mapping of charge confined in quantum dots by the scanning probe technique.
We solve the few-electron Schr\"odinger equation with the exact diagonalization approach and evaluate the energy maps 
in function of the probe position. Next, from the energy maps we try to reproduce the charge density distribution
using an integral equation given by the perturbation theory. The reproduced density maps are confronted with the original ones.
The present study covers two-dimensional quantum dots of various geometries
and profiles with the one-dimensional (1D)
quantum dot as a limit case. We concentrate on large quantum dots for which strong electron-electron correlations appear.
For circular dots the correlations lead to formation of Wigner molecules that in the presence of the tip appear in the laboratory frame.
The unperturbed rotationally-symmetric charge density is surprisingly well reproduced by the mapping.
We find in general that the size of the confined droplet as well as the spatial extent of the charge density maxima is underestimated for repulsive tip potential
and overestimated for the attractive tip. In lower-symmetry quantum dots the Wigner molecules with single-electron islands nucleate for some electron numbers even in the absence of the tip. 
These charge densities are well resolved by the mapping. The single-electron islands appear in the laboratory frame provided that classical point charge density distribution is unique,
in the 1D limit of confinement in particular.
We demonstrate that for electron systems which possess a few equivalent classical configurations
the repulsive probe switches between the configurations. In consequence the charge density evades mapping by the repulsive probe.
\end{abstract}

\maketitle

\section{Introduction}
Local properties of semiconductor nanostructures can be probed by the scanning gate microscopy  \cite{sgm} (SGM)
in which the charge of the atomic force microscope tip perturbs the potential landscape below the surface of the semiconductor within the space occupied
by confined charge carriers.
SGM covers both the open systems (quantum point contacts \cite{qpc}, resonant cavities  \cite{rc}, quantum rings  \cite{qr}) in which the probe is used to read out
the wave function at the Fermi level from  the conductance perturbations as well as systems that are weakly coupled to the reservoirs, i.e. closed quantum dots \cite{fallahi,zhang,gild07,blesz08,boyd11,boydnano,huefner11,mantelli,qiang}.
For the latter, the current flows through the quantum dot only outside the Coulomb blockade conditions  \cite{kow}, i.e., when the chemical potential ($\mu_N=E_N-E_{N-1}$) of $N$  confined electrons lies
within the transport window defined by the Fermi levels of source and drain. In experiments, the tip-induced variation of $E_N$ is determined from
the shift of voltages that is necessary for restoration of the current flow  \cite{fallahi,gild07,blesz08,huefner11}. Since the variation depends on the electron density beneath the tip this technique allows
for visualization of the charge density  \cite{scb}.
The details of the variation of the charge density can be detected only for large quantum dots exceeding the range
of tip potential. In large dots and at low electron densities the SGM technique  can -- at least potentially -- resolve single-electron islands within the few-electron quantum dot
which are formed as an effect of strong electron-electron correlations with the charge density nucleating to Wigner molecules  \cite{maksym,reiman}.

The effort on the extraction of the confined charge density by the scanning gate microscopy concerned mostly quasi one-dimensional (1D) quantum
dots defined within a quantum wire  \cite{zhang,blesz08,ziani,boyd11,boydnano,mantelli,qiang}.
Wigner molecules in 1D quantum dots  \cite{rev1d} are relatively stable against external perturbations due to the fact that the electrons in one dimension cannot exchange
their positions. In circular 2D quantum dots at low carrier densities the electron localization acquires a molecular form  \cite{maksym,reiman} only in the inner
coordinates of the system and not in the charge density which remains rotationally invariant.
 The Wigner form of the electron density with separate single-electron islands can observed in the laboratory frame of quantum dots of lower symmetry
only for some $N$, for which the classical  \cite{peeters} charge distribution reproduces the symmetry of the confinement potential  \cite{szafran}.

So far, the scanning gate microscopy experiments on closed two-dimensional (2D) quantum dots
defined within the two-dimensional electron gas (2DEG) concerned: imaging of the single-electron quantum dot  \cite{fallahi},
mapping the position of floating double  dot defined electrostatically  \cite{gild07}, and determination of the effective tip potential as
seen by the confined electrons  \cite{huefner11}.
In the present paper we consider an extent to which
the energy variation induced by the tip can be used for visualization of few-electron charge density in 2D quantum dots.
 We investigate the relation between the Wigner molecule formation in the laboratory frame and the reaction of the
 confined density on the potential of the tip. We perform configuration interaction
 calculations taking into account both the electron-electron correlation and the reaction of the quantum dot to the tip potential.
The study covers up to four electrons and quantum dots of various symmetries and profiles.
 We discuss the adequacy of the perturbative approach for extraction of the electron density  \cite{fallahi,boyd11,boydnano} confined in quantum dots
  and the fidelity of charge images outside the perturbative regime.
We find that the images obtained with repulsive (attractive) tip potential tend to overestimate (underestimate) the electron localization.
We demonstrate that the confined charge density is best resolved when the classical electron configuration agrees with the symmetry
of the confinement potential. We also identify cases when the confined electron density evades visualization
even for weak perturbation introduced by the tip.

\begin{figure}[ht!]
\hbox{
	   \epsfxsize=75mm
           \epsfbox {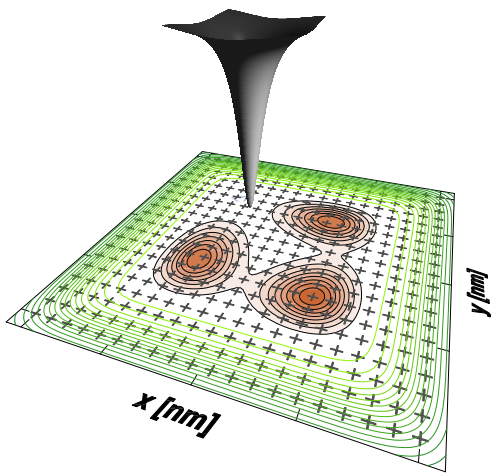}
          }
\caption{Schematic drawing of the considered system with the tip potential perturbing the charge density (brown contour plot)
of a two-dimensional quantum dot. The green contours correspond to equipotential lines. The crosses correspond to
centers of Gaussians defining the multicenter basis given by Eq. (\ref{gb}).
}
\label{schemat}
\end{figure}

\section{Theory}
In this work we assume a strictly two-dimensional model of confinement which is usually justified by the strong confinement of 2DEG in the growth direction ($z$).
The Hamiltonian of $N$-electron system is taken in form
\begin{equation}
H=\sum_i^N h({\bf r}_i) +\frac{e^2}{4\pi\epsilon\epsilon_0} \sum_{i=1}^N \sum_{j=i+1}^N \frac{1}{r_{ij}}, \label{nh}
\end{equation}
where ${\bf r}_i$ is the position of the $i$th electron and $h$ is the single-electron Hamiltonian,
\begin{equation}
h=-\frac{\hbar^2}{2m^*}\nabla^2+V({\bf r})+V_t({\bf r}; {\bf r}_t). \label{1h}
\end{equation}
 In Eq. (\ref{1h}), $m^*$ is the effective mass, $\epsilon$ is the dielectric constant (we use GaAs material parameters $m^*=0.067m_0$, $\epsilon=12.5$), $V$ stands
for the external potential, $V_t$ denotes the tip potential as seen by the electrons within the plane of confinement  and ${\bf r}_t$
is the position of the tip.

The single-electron eigenequation is diagonalized ($h\phi_n=\epsilon_n \phi_n$) using a basis of Gaussian functions
 \begin{equation}
\phi_n(x,y)=\sum_{k=1}^{M\times K} c^{(n)}_k \sqrt{\frac{2a}{\pi}} \exp\left(-a\left[(x-X_k)^2+(y-Y_k)^2\right]\right), \label{gb}
\end{equation}
where the centers of Gaussians $(X_k,Y_k)$  are distributed on a rectangular lattice of $M\times K$ points  (typically several
hundred centers are taken -- see for instance the crosses in Fig. \ref{schemat}) and $a$ is optimized variationally \cite{kkk}. The applied multicenter Gaussian basis allows for description
of any smooth potential with arbitrary or no symmetry \cite{kkk}.
The few-electron eigenproblem is solved with the configuration interaction approach using the
eigenstates of operator (\ref{1h})  for construction of the basis.
The single-electron eigenfunctions ($\phi_n$) are used for construction of Slater determinants that are used as the basis for the $N$-electron Schr\"odinger equation for Hamiltonian (\ref{nh}),
\begin{eqnarray}
&&\Psi(\left\{x_i,y_i,\sigma_i;i=1,\dots,N\right\})=\\&&\sum_k d_k {\cal{A}} \left[ \phi_{k1}(x_1,y_1)\chi_{k1}(\sigma_1)\dots\phi_{kN}(x_N,y_N)\chi_{kN}(\sigma_N)\right], \nonumber
\end{eqnarray}
where $\cal{A}$ is the antisymmetrization operator, and $\chi_k$ is one of the eigenstates of the spin Pauli $\sigma_z$ matrix.
For construction of the basis of determinants we use up to 38 single-electron spin-orbitals.

In experiment the maps of the chemical potential $\mu_N(x_t,y_t)=E_N-E_{N-1}$ can be gathered
by re-tuning the conditions for the current flow (i.e. lifting of the Coulomb blockade) with varied backgate potential or bias.
Since $\mu_1=E_1$, the energy maps as functions of the tip position can be deduced for any $N$.
Simulation of the confined charge density mapping
by the SGM technique is performed in the following sequence.
We first calculate the energy of $N-$electron system as a function of the tip position.
The charge density extracted from the energy map $n_r$ is obtained under the assumption
that the action of the tip is perturbative  \cite{fallahi,boyd11,boydnano}
\begin{equation} E_N({\bf r}_t)=E_N(\infty)+\int_{-\infty}^\infty \int_{-\infty}^{\infty}dxdy  V_t({\bf r};{\bf r}_t) n_r({\bf r}).\label{feft} \end{equation}
Equation (\ref{feft}) is a convolution of the tip potential and the confined charge density.
The charge density can be extracted from the energy dependence of the tip position using the Fourier transform technique  \cite{fallahi,boyd11}.
Alternatively Eq. (\ref{feft}) can be treated as the Fredholm-type integral equation for $n_r$. We apply the Nystr\"om approach replacing
the integral by a quadrature (rectangle rule). Upon replacement we obtain a linear system of equations for the charge density in the mesh points used for the quadrature.
In the following we compare the charge density deduced in this way  ($n_r$) with the exact one ($n$)
that is calculated from the few-electron wave function
\begin{equation}
n({\bf r})=\langle \Psi|\sum_{i=1}^N \delta({\bf r}-{\bf r_i})|\Psi \rangle.
\end{equation}
For comparison we  calculate also the charge density $n_\delta$ assuming that the tip potential
is point-like $V_p=V_T\delta(x-x_t,y-y_t)$. Then, Eq. (\ref{feft}) reduces to
\begin{equation} E_N({\bf r}_t)=E_N(\infty)+V_T n_\delta({\bf r}_t).\label{fefut} \end{equation}
Normalized charge density $n_\delta$ derived from this formula is simply proportional to the
energy variation.

The original potential of the charged tip is of the Coulomb form. As the tip approaches the 2DEG its charge density reacts
by deformation which results in the screening of the tip potential. Previous Schr\"odinger-Poisson calculations \cite{szafranr} for the reaction of the 2DEG
to the tip indicated that the effective (i.e. screened) tip potential is close to the Lorentz form
\begin{equation}
V_t({\bf r};{\bf r}_t)=\frac{V_T}{\frac{(x-x_t)^2}{d_{tip}^2}+\frac{(y-y_t)^2}{d_{tip}^2}+1},
\end{equation}
where the width of the tip $d_{tip}$ turns out to be of the order of the tip -- 2DEG distance \cite{szafranr} independent of the charge at the tip and the density of 2DEG, which only
influence the strength of the perturbation $V_T$ and not its range. The Lorentz form of the effective tip potential for SGM measurements on 2DEG was also found in experimental studies of
2D quantum dots  \cite{huefner11}.
Therefore, in the bulk of this work we use mainly the Lorentz model potential of the tip and assume $d_{tip}=20$ nm, which seems the smallest reasonable value
of the potential width. For comparison we also consider the reaction of the confined system to the long-range Coulomb potential
\begin{equation}
V^c_t({\bf r};{\bf r}_t)=\frac{Qe}{4\pi\epsilon\epsilon_0} \frac{1}{\left((x-x_t)^2+(y-y_t)^2+z_{tip}^2\right)^{1/2}},
\end{equation}
where $Q$ is the tip potential, and $z_{tip}$ -- the position of the tip above the 2DEG plane.

An ample discussion of potentials of electrostatic quantum dots defined within 2DEG was given in Ref.  \cite{lis}.
The calculations  \cite{lis} indicated that depending on the geometry of the device one can obtain both parabolic and
quantum well profiles. The latter can only be realized for large dots \cite{lis} of the linear extent of the order of a few hundred nm, i.e. the ones which are studied in the present work.
Since our purpose is to determine the relation between the $N-$dependent classical electron distribution
and the quantum dot geometry we consider a number of potentials: parabolic
\begin{equation}
V(x,y)=\frac{m}{2} \left(\omega_x^2 x^2 +\omega_y^2 y^2\right),
\end{equation}
of circular $\omega_x=\omega_y=\omega$ and elliptic $\omega_x\neq\omega_y$ symmetry,
as well as non-parabolic dots modeled by formula
\begin{equation}
V(x,y)=V_0\left(1-\frac{1}{1+(\frac{x}{X})^{10}+(\frac{y}{Y})^{10}}\right) \label{rect}
\end{equation}
which produces a well-like potential with a flat bottom of a nearly rectangular shape
of dimensions $2X\times 2Y$. For Eq. (\ref{rect}) we discuss the dots
from a nearly 1D ($X>>Y$) to square ($X=Y$) geometry.

\begin{figure*}[ht!]
\hbox{
	   \epsfxsize=130mm
           \epsfbox {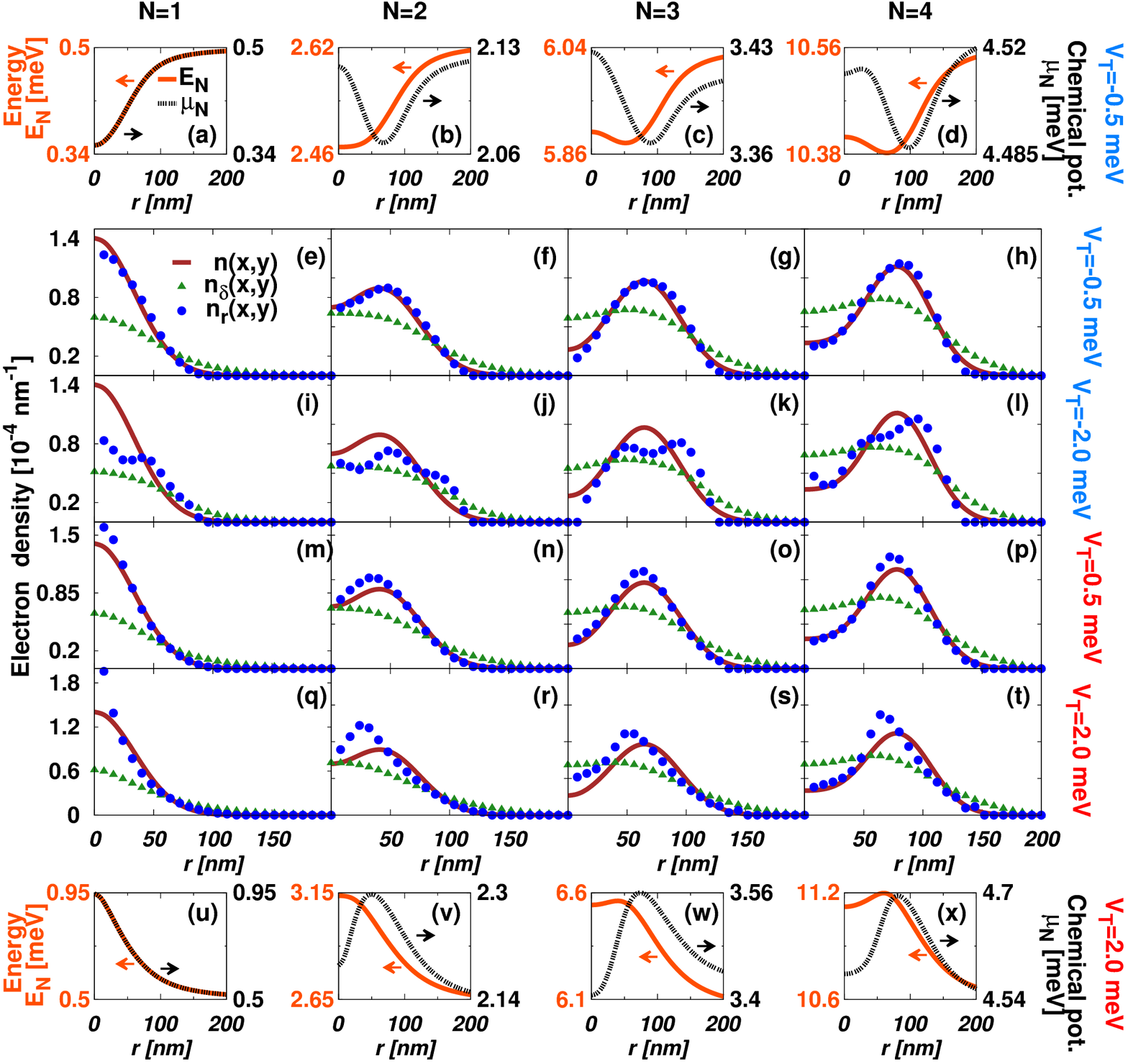}
          }
\caption{Results for a circular harmonic oscillator potential with $\hbar \omega=0.5$ meV and the potential
of the tip of the Lorentz form [Eq.(8) with $d_{tip}=20$ nm].
The columns correspond to various electron number from $N=1$ to 4.
The first (last) row of plots shows the energies and chemical potentials for $V_T=-0.5$ meV ($V_T=2$ meV).
The rows from the second (e-h) to the fifth (q-t) show the electron densities: the unperturbed one ($n$ brown solid
lines), the one reproduced with the perturbative formula (5) ($n_r$, blue dots) and the one obtained
under assumption of a delta-like perturbation ($n_\delta$, green triangles).
Subsequent rows correspond to various values of $V_T$.
}
\label{oscylator}
\end{figure*}

\begin{figure}[ht!]
\hbox{
	   \epsfxsize=70mm
           \epsfbox {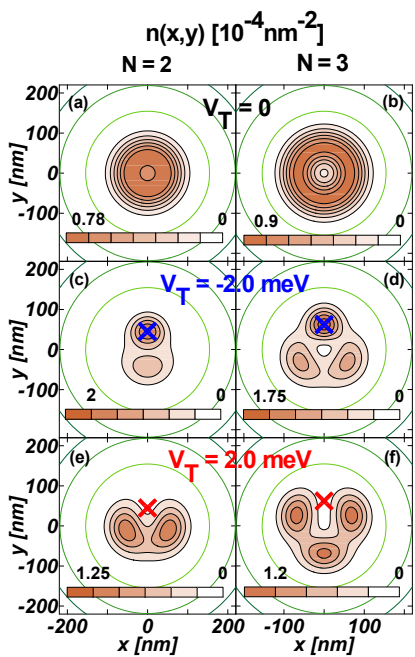}
          }
\caption{Charge densities for $N=2$ and 3 electrons for parameters of Fig. \ref{oscylator}. The cross
marks the tip position. The solid green curves show the equipotential lines.
}
\label{oscylatorg}
\end{figure}

\begin{figure}[ht!]
\hbox{
	   \epsfxsize=100mm
           \epsfbox {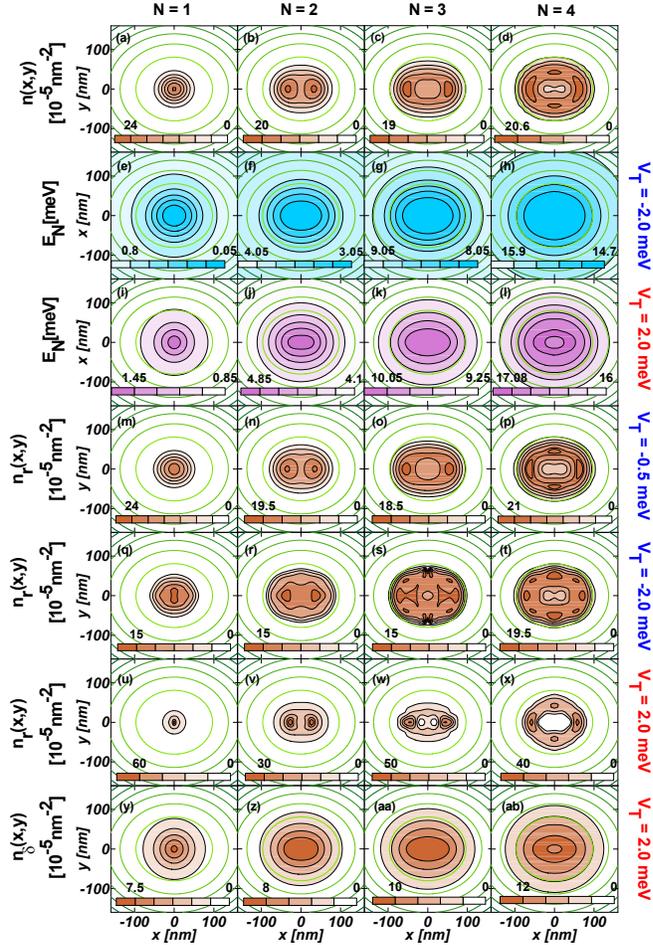}
          }
\caption{Results for elliptic parabolic potential with $\hbar \omega_x=0.8$ meV,  $\omega_y=1.2\omega_x $, and the potential
of the tip of the Lorentz form [Eq.(8) with $d_{tip}=20$ nm].
Columns correspond to various electron numbers. The first row of plots (a-d) shows
the charge density in the absence of the tip. Second (e-h) and third (i-l) rows show the energies
in function of the tip position for $V_t=-2$ meV and $V_t=2$ meV, respectively.
Three next rows show the charge density reproduced by the perturbative formula, and the last
one the density calculated for the assumption of the point-like tip potential.}
\label{eliptu}
\end{figure}

\begin{figure}[ht!]
\hbox{
	   \epsfxsize=100mm
           \epsfbox {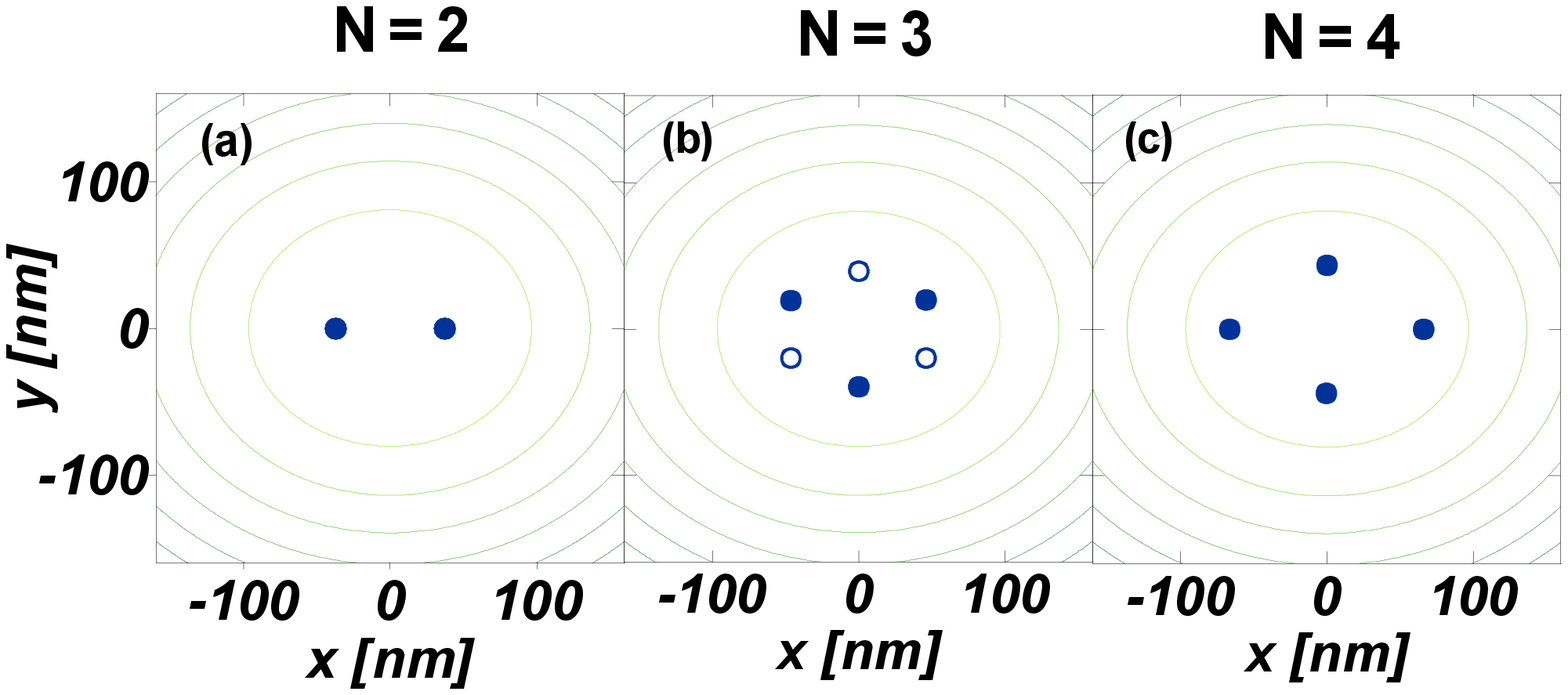}
          }
\caption{Classical lowest energy configurations of point charges for the elliptic potential of Fig. \ref{eliptu}.
For $N=3$ two equivalent configurations exist which are marked by solid and empty circles, respectively.}
\label{celiptu}
\end{figure}

\section{Results and Discussion}

\subsection{Circular potential}
Let us first consider a circular parabolic quantum dot with $\hbar \omega=0.5$ meV.
The solid line in Fig. \ref{oscylator}(e,f,g,h) shows the unperturbed radial charge density for 1,2,3 and 4 electrons,
respectively. When a tip modeled by the attractive Lorentzian with $V_t=-0.5$ meV moves above the system,
the changes in the energy are of the order of 0.2 meV only [Fig. \ref{oscylator}(a-d)].
The reproduced charge density: $n_r$ (circles) very well agrees with the unperturbed density,
which is also the case for weak repulsive tip potential [Fig. \ref{oscylator}(m-p)].
A stronger amplitude of the energy variation should be useful for the signal to noise ratio
of the experimental maps. For a stronger attractive perturbation ($V_t=-2$ meV, Fig. \ref{oscylator}(i-l)) we notice that the
size of the charge droplet is overestimated with an extra oscillation of a small amplitude.
On the other hand: a stronger repulsive potential ($V_t=2$ meV, Fig. \ref{oscylator}(q-t)) gives a smaller size of the droplet.

Figure \ref{oscylatorg} shows the two- and three- electron densities: unperturbed [Fig. \ref{oscylatorg}(a-b)],
and in the presence of the tip [Fig. 3(c-f)]. The attractive tip $V_t=-2$ meV captures a part of the density underneath [Fig. \ref{oscylatorg}(c-d)],
the position of the other electron islands becomes well resolved.  The repulsive tip also pins the orientation
of the Wigner molecule in the laboratory frame [Fig. \ref{oscylatorg}(e-f)], only with a void under the tip position.
The impact of the tip on the electron density is therefore a drastic one for both negative and positive tip potentials, so a success of the perturbative
formula given by Eq. (\ref{feft}) for reproduction of the radial density as calculated in the absence
of the tip found in Fig. \ref{oscylator} is quite remarkable, even if the original density lacks finer details.
\subsection{Elliptic potential}

The finer details of the unperturbed electron density appear for dots of lowered symmetry:
see the results for an elliptic ($\hbar \omega_x=0.8$ meV, $\omega_y=1.2 \omega_x$) quantum dot in Fig. \ref{eliptu}.
For two electrons: single-electron islands are formed [Fig. \ref{eliptu}(b)] along the $x$ axis, as in the classical solution
[Fig. \ref{celiptu}(a)].  For three electrons
two equivalent classical charge distributions exists [see Fig. \ref{celiptu}(b)] each of the symmetry lower than the elliptical one, and in consequence
the single-electron islands do not appear in the laboratory frame of the quantum system [Fig. \ref{eliptu}(c)].
For each $N$, similarly as for the circular potential, for a weak perturbation both negative tip [$V_t=-0.5$ meV -- Fig. \ref{eliptu}(m-p)]
and positive tip [not shown], $n_r$  density very well reproduces the unperturbed one.

Second and third rows of Fig. \ref{eliptu} show the energy of the system in function of the tip position for stronger
attractive and repulsive tip potentials $V_T=\pm 2$ meV. For the negative tip we obtain a flat minimum, within
more or less the entire region occupied by the unperturbed charge density [Fig. \ref{eliptu}(e-h)]. The minimum appears since the tip creates its own potential
minimum and the electron system follows the minimum as the tip is translated.  The reproduced density [Fig. \ref{eliptu}(q-t)] occupies
larger space than the original one (as in the circular for $V_t=-2$ meV -- see Fig. \ref{oscylator}(i-l)), and a variation of smaller amplitude appears
within the maximal density area.

For the positive tip the energy maximum is more strongly localized [Fig. \ref{eliptu} (i-l)] around the center of the dot,
and the amplitude of $n_r$ variation [Fig. \ref{eliptu} (u-x)] is drastically increased with respect to $n$.
Also, the reproduced density occupies visibly smaller area than the original one. Nevertheless,
the configuration of the maxima of $n_r$ agrees with the ones present for $n$.

For $V_T=2$ meV, the density $n_\delta$ reproduced with the assumption that the energy map --
obtained in fact for the Lorentz function -- is due to the delta-like tip potential, gives a closer [Fig. \ref{eliptu} (y-ab)]
idea about the size of the charge droplet. Nevertheless, the details of the density maxima are not reproduced by $n_\delta$.

\subsection{Square quantum dot}

\begin{figure}[ht!]
\hbox{
	   \epsfxsize=100mm
           \epsfbox {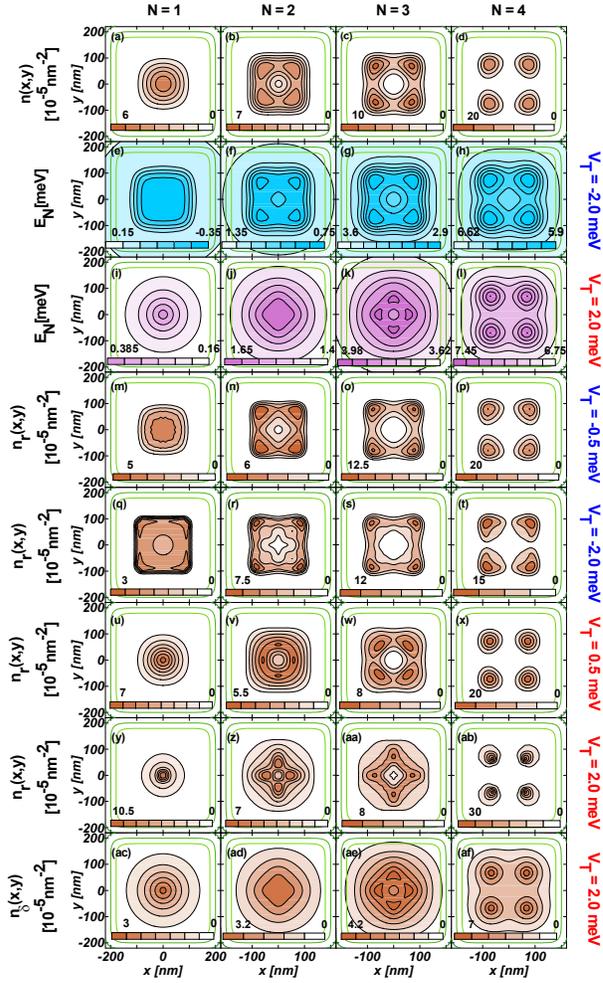}
          }
\caption{Results for a square quantum dot $X=Y=200$ nm [Eq. (11)] and the tip potential
the Lorentz form [Eq.(8) with $d_{tip}=20$ nm].
Columns correspond to various electron numbers. The first row of plots (a-d) shows
the charge density in the absence of the tip. Second (e-h) and third (i-l) rows show the energies
in function of the tip position for $V_t=-2$ meV and $V_t=2$ meV, respectively.
Three next rows show the charge density reproduced by the perturbative formula, and the last
one -- the density calculated for the assumption of the point-like tip potential.}
\label{square}
\end{figure}

\begin{figure}[ht!]
\hbox{
	   \epsfxsize=100mm
           \epsfbox {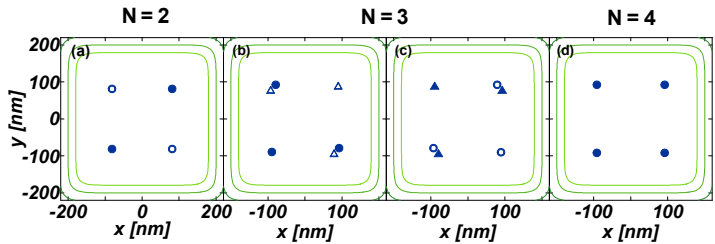}
          }
\caption{Classical lowest energy configurations of point charges for the potential of Fig. \ref{square}.
For $N=2$ (a) two equivalent configurations exist which are marked by different symbols.
For $N=3$ four equivalent configurations appear (b,c).}
\label{csquare}
\end{figure}

\begin{figure}[ht!]
\hbox{
	   \epsfxsize=70mm
           \epsfbox {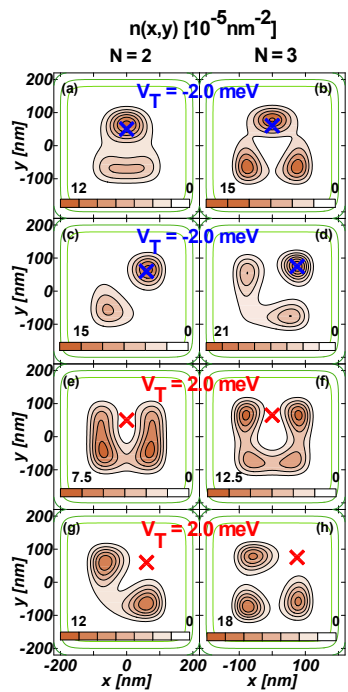}
          }
\caption{Charge densities for $N=2$ and 3 electrons for parameters of Fig. \ref{square}. The cross
marks the tip position.}
\label{squareg}
\end{figure}

\begin{figure}[ht!]
\hbox{
	   \epsfxsize=100mm
           \epsfbox {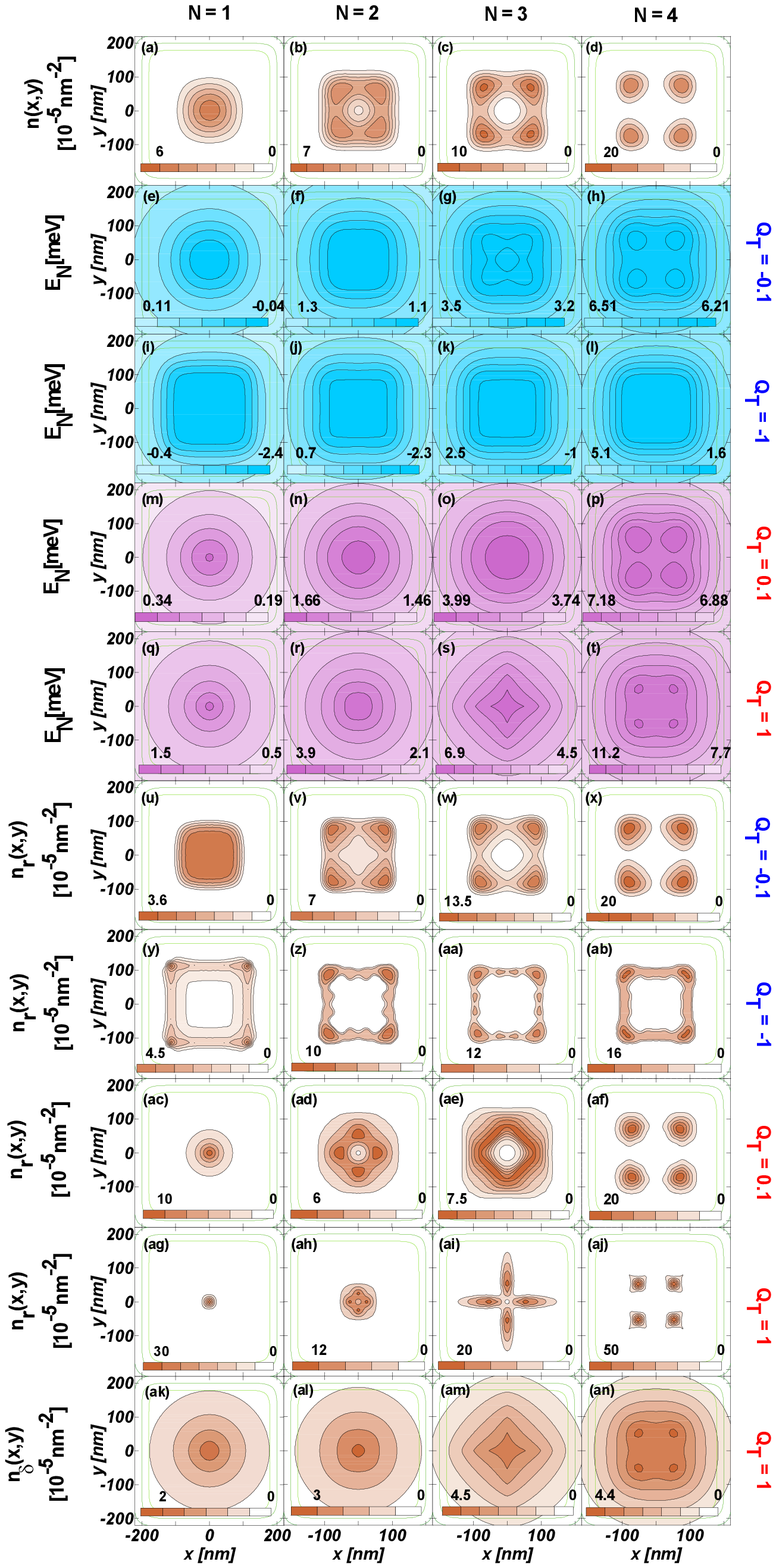}
          }
\caption{Results for the square quantum dot $X=Y=200$ nm [Eq. (11)] and the tip potential
the Coulomb form [Eq.(9) with $z_{tip}=30$ nm, $Q=\pm 0.1e$ and $Q=\pm 1e$].
Columns correspond to various electron numbers. The first row of plots (a-d) shows
the charge density in the absence of the tip. Plots (e-t) show the energies
as functions of the tip position.
Four next rows show the charge density reproduced by the perturbative formula, and the last
one -- the density calculated for the assumption of the point-like tip potential.}
\label{squarec}
\end{figure}

Let us now consider a square quantum dot [confinement potential given by Eq. (11) for $X=Y=200$ nm].
For $N>1$ the unperturbed electron density distinctly sticks to the corners of the square dot [Fig. \ref{square}(b-d)]
as should be expected for interacting charge density within a box.
For $N=2$ and 3 an increase of the density along the dot edges is also observed.
The classical system of 4 electrons possesses a single lowest-energy configuration [Fig. \ref{csquare}(d)] which coincides
with the charge density of Fig. \ref{square}(d).  Systems of two and three electrons possess
two and four equivalent configurations, respectively [Fig. \ref{csquare}(a-c)]. The few-electron
wave function contains contributions of all these configurations. Since one or two corners
of the square are unoccupied in the classical configurations, the lower amplitude of the maxima at the corners
than in the case of $N=4$.
As the negative tip scans the dot [Fig. \ref{square}(e-h)] we observe a rather flat dependence of the energy on
the tip position with minima at the corners -- where the unperturbed electron density is the largest.
Remarkably, for the positive tip and $N=2$ and $N=3$ the energy extrema lie on the axes of the dot [Fig. \ref{square}(j-k)], and not on
the corners. The reason for this behavior is given in Fig.\ref{squareg} which displays the charge distribution
when the tip is present. When the tip is above the corner of the dot [Fig. \ref{squareg}(g-h)] we can see that
the electron density becomes nearly identical with the classical systems of electrons [cf. Fig. \ref{csquare}(a-c)].
The other equivalent lowest-energy configurations of the charge density were excluded by the presence
of the tip above one of the corners. On the other hand, for two-electrons when the tip is above the center of the side of the dot [Fig. \ref{squareg}(e)]
the electrons instead of occupying the opposite corners of the dot, go to  its other side and approach each other, hence
the corresponding maximum of the energy [Fig. \ref{square}(j)]. For three electrons, formation of four instead
of three charge maxima is found [Fig. \ref{squareg}(f)] which also increases the electrostatic energy of the system above
the minimal one.

The charge density $n_r$ as reproduced for the negative tip [Fig. \ref{square}(m-t)] exhibits maxima near the maxima of $n$.
For stronger tip potential [Fig. \ref{square}(q-t)] the charge density localization near the potential edges is overestimated.
When a weak positive tip potential ($V_T=0.5$ meV) is applied [Fig. \ref{square}(u-v)] we notice that the maxima are pushed to the interior of the dot from the edges.
We also notice for two-electrons [Fig. \ref{square}(v)] that the position of the $n_r$ maxima is shifted to the axes of the dot,
which is the result of the energy increase for the tip above the sides of the dot discussed above.  For $V_T=2$ meV
we notice a similar phenomenon also for three electrons [Fig. \ref{square}(aa)]. For one and four electrons [Fig. \ref{square}(y,ab)]
the maxima are localized in the correct positions, and they distinctly shrink in size as compared to the maxima of $n$.

We also considered the mapping of the charge density confined in the square quantum dot by the long-range Coulomb potential [Eq. (9)]
i.e. for neglected screening of tip potential -- see Fig. \ref{squarec}. We considered the tip localized $z_{tip}=30$ nm above the dot and the charge at the tip $Q=\pm 1$ and $\pm 0.1$ [e],
for which the maximal value of the perturbation below the tip equals 3.8 meV and 0.38 meV, respectively.
The weak ($Q=-0.1$) attractive perturbation [Fig. \ref{squarec}(u-x)] gives $n_r$ which well agree with $n$.
On the other hand already for the weak repulsive perturbation ($Q=0.1$) the maxima of $n_r$ for $N=2$ and 3 go to the axes of the dot --
the phenomenon observed above for the Lorentz perturbation. For $Q=\pm 1$ the calculated $n_r$
differs drastically from the unperturbed density $n$. For $Q=-1$,  $n_r$ [Fig. \ref{squarec}(y-ab)] drifts to the edges of the dot.
This density localization convolved [Eq. (5)] with the Coulomb potential gives the flat energy curve of Fig. \ref{squarec}(i-l).
On the other hand for the positive potential $Q=1$ the derived $n_r$ density [Fig.\ref{squarec}(ag-aj)] is localized in tiny islands inside the dot
and only for a single and four electrons -- for which a single classical configuration of the charge exists -- their positions are close to the original ones.
For $N=2$ and $N=3$ the switching between similar configurations of the type presented in Fig. \ref{squareg} becomes so strong, that the reproduced charge density [Fig. \ref{squarec}(ah,ai)] have little in common
with the original one.

\subsection{Rectangular quantum dot near 1D limit}
\begin{figure}[ht!]
\hbox{
	   \epsfxsize=100mm
           \epsfbox {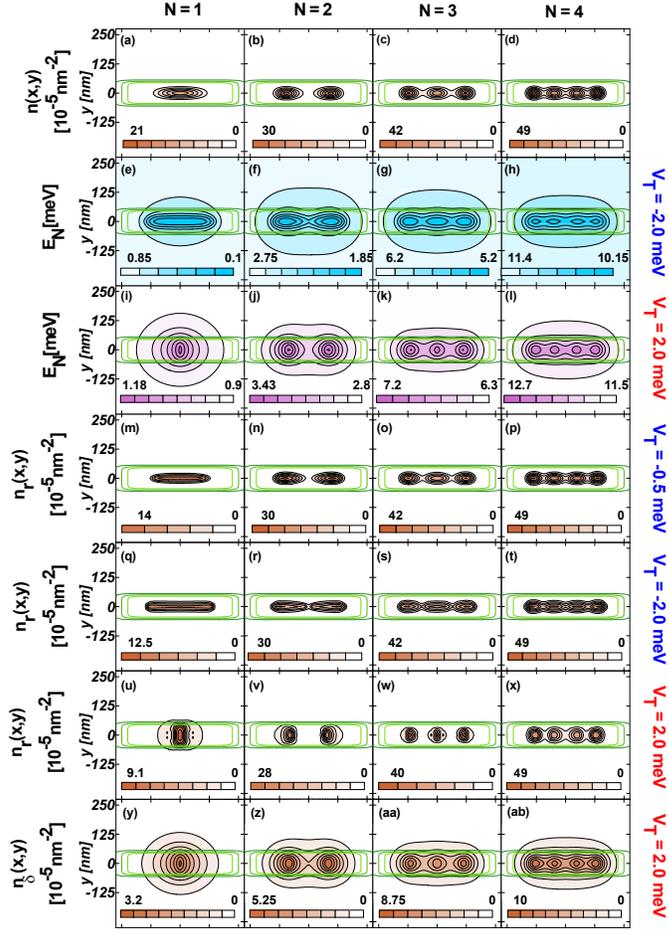}
          }
\caption{Results for the rectangular quantum dot $X=250$nm,  $Y=50$ nm [Eq. (11)] and the tip potential
the Lorentz form [Eq.(8) with $d_{tip}=20$ nm].
Columns correspond to various electron numbers. The first row of plots (a-d) shows
the charge density in the absence of the tip. Second (e-h) and third (i-l) rows of plots show the energies
as functions of the tip position.
Three next rows show the charge density reproduced by the perturbative formula, and the last
one -- the density calculated for the assumption of the point-like tip potential.}
\label{1dl}
\end{figure}

\begin{figure}[ht!]
\hbox{
	   \epsfxsize=100mm
           \epsfbox {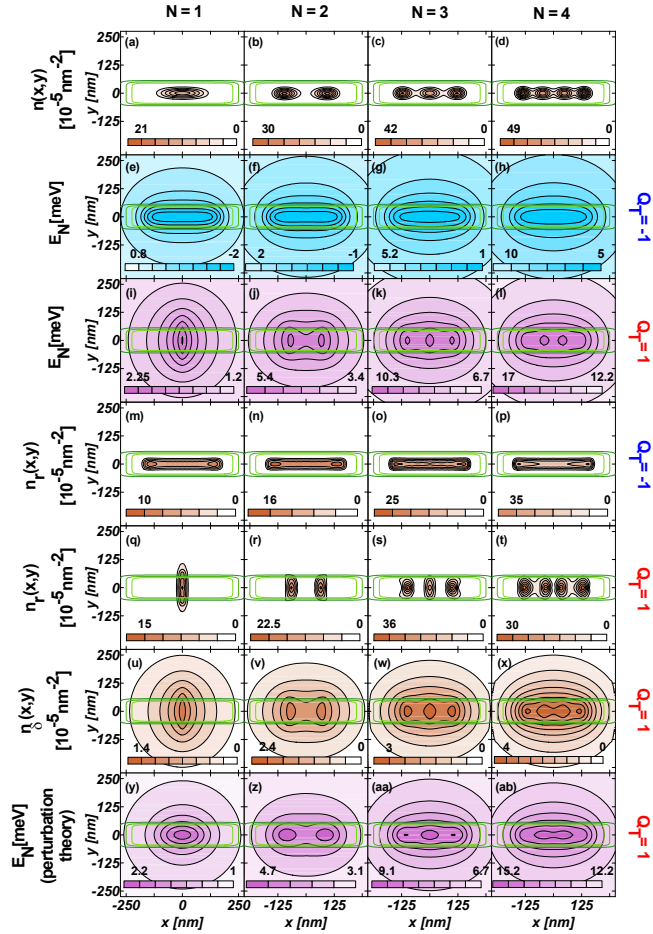}
          }
\caption{Same as Fig. \ref{1dl} only for the tip potential of
the Coulomb form [Eq.(9)] with $z_{tip}=30$ nm, $Q=\pm 0.1$ [e] and $Q=\pm 1$ [e].}
\label{1dc}
\end{figure}

\begin{figure}[ht!]
\hbox{
	   \epsfxsize=100mm
           \epsfbox {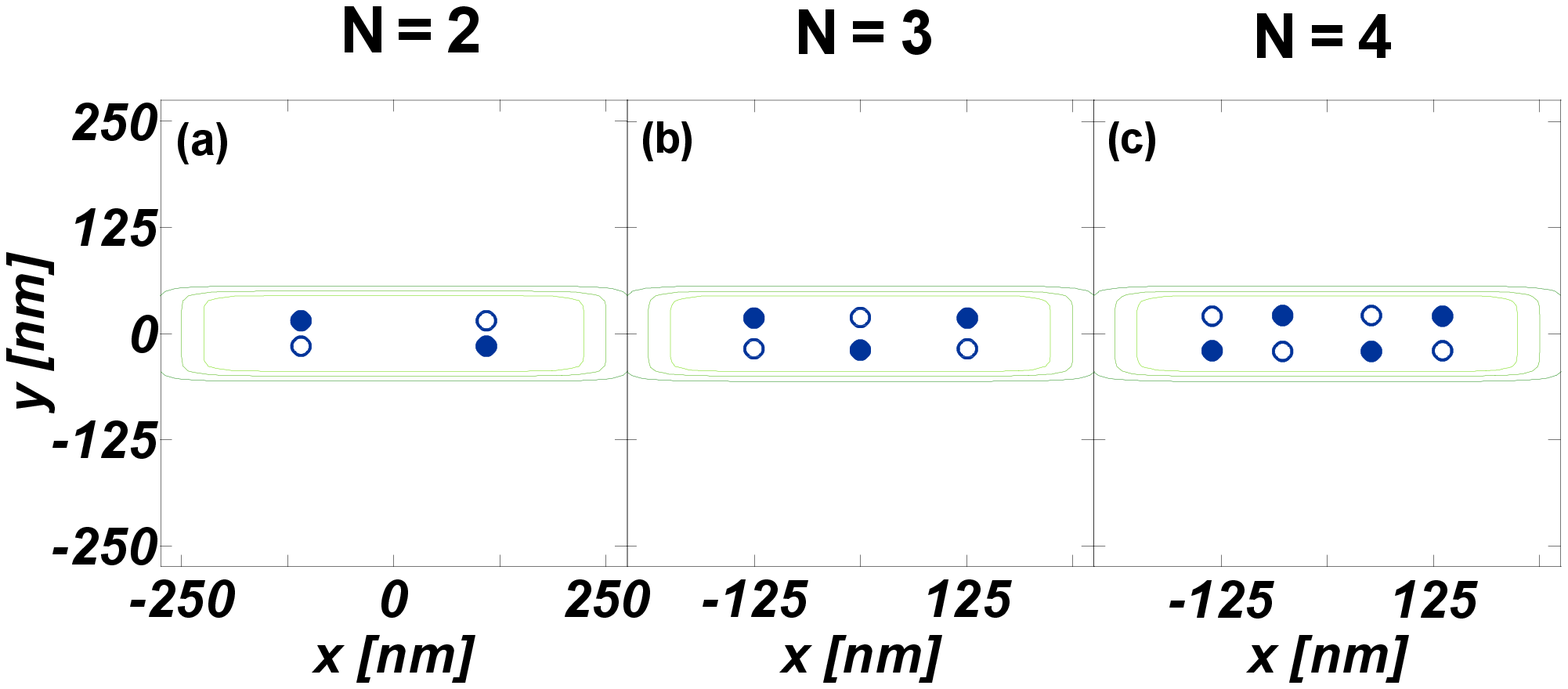}
          }
\caption{Classical lowest energy configurations of point charges for the potential of Fig. \ref{1dl}.
For each $N$ considered two equivalent configurations appear.}
\label{c1d}
\end{figure}

Finally, let us consider the rectangular quantum dot near the 1D limit ($X=250$ nm, $Y=50$ nm).
The results for the Lorentz and Coulomb tip potentials are displayed in Figs.\ref{1dl} and \ref{1dc}, respectively.
For this dot the classical few-electron systems for $N=2,3,4$ possess two equivalent zig-zag
configurations [Fig.\ref{c1d}]. The electron positions in the two configurations differ only slightly
and these differences are not resolved in the quantum charge density,
which  presents well resolved single-electron islands that are clearly visible in the original electron density for all $N$ [Fig. \ref{1dl}(a-d)].
The single-electron islands are also well resolved in the maps obtained for all the Lorentz tip potentials considered in Fig. \ref{1dl}(m-x),
as well as in the crudest assumption of point-like potential [Fig. \ref{1dl}(y-ab)].
The charge density islands obtained for $V_T=-2$ meV [Fig. \ref{1dl}(q-t)] are distinctly more
extended along the axis of the dot, which results from the fact that the islands
follow the attractive tip as it moves. On the other hand for $V_T=2$ meV [Fig. \ref{1dl}(u-x)]
an extension of the $n_r$ densities in the direction perpendicular to the axis is found.
When the repulsive tip  localized near the edges of the dot its width in $y$ direction
is effectively reduced. In this way the tip increases not only
the local potential energy but also the kinetic energy due to the localization.
For that reason, the perturbative formula produces the charge density which penetrates the region
outside the dot, where the original density vanishes.

The repulsive Coulomb potential reproduces correctly the charge localization [Fig. \ref{1dc}(q-t)]
with an enhanced effect of the elongation of charge density island perpendicular to the dot.
On the other hand, the attractive Coulomb potential [Fig. \ref{1dc}(m-p)] misses the details
of the charge density which is seen as equally spread along the dot.
For $Q=\pm 0.1$  (not shown) the 1D Wigner molecule is well reproduced by $n_r$.

\section{Summary and Conclusions}

We have performed simulations of the charge density mapping for electron systems confined in two-dimensional quantum dots
using model tip potentials of the Lorentz and Coulomb form, several confinement potentials and the exact solution of the few-electron Schr\"odinger equation.
We investigated large quantum dots, where the electron-electron correlation is strong, which can give rise to formation of single-electron islands
in the laboratory frame, i.e. the Wigner molecules.

For the circular dots we found that the molecular electron distributions appear in the laboratory frame
pinned by the tip potential. The Wigner molecule follows the tip as it moves above the dot. In consequence
the energy map in function of the tip position is rotationally invariant, and the density map reproduced
by the perturbative formula is very close to the original one. We noticed that a stronger repulsive (attractive) tip
leads to underestimate (overestimate) of the charge density size.
This conclusions for the size of the droplet holds for any dot profile studied.
Moreover, for the elliptical and square quantum dots single-electron islands appear in the charge density
for some $N$, and they are resolved in the charge density that is reproduced from the integral perturbative formula.
We have found that in the 1D dots the Wigner molecule is clearly visible in the charge density mapped from the
energy dependence for all $N$ and for most of the tip potential studied, with the exception of the negative Coulomb
potential for which the single-electron islands are lost.
We demonstrated that the charge density of electron systems which possess a few
equivalent classical configurations, generally are difficult to be resolved by the scanning probe technique
for the repulsive tip potential, since the tip switches between equivalent configurations. In consequence, the mapped charge density  maxima
do not overlap with the original ones.

\section*{Acknowledgments}
This work was supported by National Science Centre
according to decision DEC-2012/05/B/ST3/03290 and by
PL-Grid Infrastructure. Calculations were performed in ACK--
CYFRONET--AGH on the RackServer Zeus.


\begin{thebibliography}{99}
\bibitem{scb} This technique is called Coulomb blockade microscopy by some groups \cite{mantelli,qiang}.
\end{thebibliography}
\end{document}